\documentclass[2p,11pt]{elsarticle}

\usepackage{graphicx}
\usepackage{xcolor}
\usepackage{diagbox}
\usepackage{amsmath,amssymb,amsthm}
\usepackage[english]{babel}
\usepackage[utf8]{inputenc}
\usepackage{hyperref}

\definecolor{dblue}{RGB}{0,0,90}
\definecolor{gris}{RGB}{0,20,20}

\usepackage{hyperref} 
\hypersetup{hidelinks,colorlinks,breaklinks=true,urlcolor=dblue,citecolor=dblue,
	linkcolor=dblue,bookmarksopen=false}

\journal{The International Journal of Biostatistics}

\begin{document}

\begin{frontmatter}



\title{Multinomial logistic model for coinfection diagnosis
between arbovirus and malaria in Kedougou}

\author{Mor Absa Loum $^{1\ast}$}
\author{Marie-Anne POURSAT $^1$}
\author{Abdourahmane SOW $ ^{2}$}
\author{Amadou Alpha SALL $ ^{2}$}
\author{Cheikh LOUCOUBAR $ ^{3}$}
\author{Elisabeth Gassiat $^1$}

\cortext[cor]{Corresponding author: mor-absa.loum@u-psud.fr}

\address{$ ^{1} $ Laboratoire de Mathématiques d'Orsay, Univ. Paris-Sud, CNRS, Universit\'e Paris-Saclay, 91405 Orsay, FRANCE\\ $ ^{2} $Institut Pasteur de Dakar, Arboviruses and Viral Hemorrhagic Fevers Unit \\  $ ^{3} $Institut Pasteur de Dakar, Biostatistics, Bioinformatics and Modeling Group}

\begin{abstract}
In tropical regions, populations continue to suffer morbidity and mortality from malaria and arboviral diseases. In Kedougou (Senegal), these illnesses are all endemic due to the climate and its geographical position. The co-circulation of malaria parasites and arboviruses can explain the observation of coinfected cases. Indeed there is strong resemblance in symptoms between these diseases making problematic targeted medical care of coinfected cases. This is due to the fact that the origin of illness is not obviously known. Some cases could be immunized against one or the other of the pathogens, immunity typically acquired with factors like age and exposure as usual for endemic area. Thus, coinfection needs to be better diagnosed. Using data collected from patients in Kedougou region, from $2009$ to $2013,$ we adjusted a multinomial logistic model and selected relevant variables in explaining coinfection status. We observed specific sets of variables explaining each of the diseases exclusively and the coinfection. We tested the independence between arboviral and malaria infections  and derived coinfection probabilities from the model fitting. In case of a coinfection probability greater than a threshold value to be calibrated on the data, long duration of illness and age are mostly indicative of arboviral disease while high body temperature and presence of nausea or vomiting symptoms during the rainy season are mostly indicative of malaria disease. 
\end{abstract}

\begin{keyword}
Arbovirus\sep  coinfection\sep malaria\sep multinomial logistic regression\sep random forest classification\sep variable selection.
\end{keyword}

\end{frontmatter}



\section{Introduction}

\label{sec1}

Concurrent infections are often observed among vector borne diseases such as malaria and arthropod-borne viral diseases (arbovirus) in tropical regions (\cite{Mohapatra}, \cite{Mushtaq}). It is the case for malaria and dengue in American, African and Asian tropical regions where their endemic areas overlap largely (\cite{Carme,Arya,Deresinski,Ali,Senn,Charrel,VRRM}). Malaria can be easily ascribed to other febrile illnesses because its clinical symptoms are often indistinguishable from those initially seen in dengue or chikungunya for instance (\cite{Baba}). Since the introduction of the Rapid Diagnostic Test (RDT) in $2007$ in Senegal, malaria has been better diagnosed and an important decrease has been noticed in the prevalence of malaria. Thus we may think that malaria has been overestimated for some time at the expense of other febrile diseases such as arbovirus or bacteria (\cite{Thiam,Prog}). Presumptive treatment of fever with antimalarial is widely practiced to reduce malaria attributable mortality. This practice means that ill patients may be inappropriately treated, particularly where rapid diagnosis test kits are not readily available, or if the opportunity to test for arboviral infections is missed. Thus, misdiagnosis of arbovirus coinfections as malaria infections may be a cause for underestimating emerging arbovirus infections. In $2009$, surveillance of acute febrile illness (AFI) was implemented in Kedougou for early detection of arbovirus outbreaks and malaria and in order to accurately measure disease morbidity and mortality in this geographical region. Due to co-circulation of malaria parasites and arbovirus, that were mainly dengue (DEN), chikungunya (CHIK), Zika (ZIK), yellow fever (YF) and Rift Valley fever viruses (RVFV) in this region (neglecting the prevalence of other arboviral infections), concurrent infections were observed and posed a challenge for medical diagnosis (\cite{sow}). 
Here we compare clinical profiles of coinfected patients to clinical profiles of mono-infected patients through the statistical analysis of a data set collected from febrile patients in the Kedougou region, Senegal from $2009$ to $2013$. Our study aims to characterize the risk factors of coinfection and to provide statistical indicators that improve differential diagnosis of febrile cases for arbovirus.

\smallskip
The data of our study were provided by the Institut Pasteur de Dakar (IPD) at Kedougou (southern-east Senegal). In this region, malaria and arbovirus are endemic due to the climate and the population movements. Data were collected through seven healthcare centers in the region: \textit{Ninefesha rural hospita}l, \textit{Kedougou} and \textit{Saraya Health Centers, Bandafassi} and \textit{Khossanto health posts}, \textit{the Kedougou military health post,} and the \textit{Catholic Mission mobile team}. Inclusion criteria were (i) being at least one year old at the date of the visit, (ii) having fever (i.e., body temperature$\geq 38^{o}C$) and (iii) manifesting at least one clinical sign within a list of symptoms. Patients satisfying inclusion criteria were enrolled once a written informed consent was signed.

In the present paper, we propose a multinomial logistic model to analyse coinfection between arbovirus and malaria. There were four outcomes determining four groups of patients: arbovirus monoinfections (with respect to the 5 tested arbovirus), malaria monoinfections, arbovirus-malaria coinfections and controls defined as patients negative for malaria and for the $5$ tested arbovirus. Febrile episodes from this control group were probably due to other circulating pathogens for which all groups were supposed to be equally exposed. We first performed a covariable selection using random forests based on the variable importance measure (\cite{Genuer1}). Secondly we fitted a parametric multinomial logistic model including the selected covariables and  quantified the influent factors on the different outcomes to investigate the following questions: Which factors can explain coinfection? Which risk factors enable to distinguish between malaria and arbovirus?  Finaly, we proposed a Wald-type test to test the correlation between malaria infection and arboviral infection.  If the independence hypothesis is rejected, we were able to predict the probability that a patient be coinfected given that malaria is observed. This predictive analysis was illustrated on simulated data.

%


The paper is organized as follows. 
In Section \ref{sec2}, we present the working data set. Section \ref{sec3} describes the statistical model and the variable selection. In Section \ref{sec4}, we present the independence test between arbovirus and malaria infections and we propose a predictive analysis. A concluding discussion is given in Section \ref{sec5}. Additional analysis and results are provided in Supplementary Material.

\section{Data description}
\label{sec2}
We based our analysis on the data from IPD at Kedougou. The initial data set included $15\;523$ patients and collected various features: patients' data (like sex, age, occupation, location,$\ldots$),  clinical symptoms, climate indicators and three binary infections status variables indicating $(i)$ the presence or absence of malaria parasites in blood, $(ii)$ the detection of virus or IgM antibodies against virus. Malaria diagnosis relied on the identification of haematozoa using the thick blood smear (TBS) method. Arboviral infections were investigated by the detection of specific anti-arbovirus IgM using ELISA (enzyme-linked immunosorbent assay). We considered an \emph{arboviral} case as any individual tested positive to the infection with at least one of the five arbovirus (DEN, CHIK, ZIK, YF and RVF).

Based on these data we created a new categorical response variable built from four possible combinations of 
the three infection status variables as follows:
$$ Y = \left\{
\begin{array}{ cl}
0 & \mbox{``Other febrile illnesses (O)"} \\
1 & \mbox{``Arboviral monoinfection (A)"}  \\
2 & \mbox{``Malaria monoinfection (M)"}  \\
3 & \mbox{``Coinfection (C)"}
\end{array}\right.
$$
Category $0$ corresponds to individuals that are negative for both malaria and the tested arboviral infections; their symptoms could be due to other unknown febrile illnesses. 
Category $1$ corresponds to individuals positive for at least one of the five tested arbovirus and negative for malaria.
Category $2$ corresponds to individuals negative for tested arbovirus and positive for malaria.
Category $3$ represents individuals simultaneously positive for malaria and for at least one of the tested arbovirus. The subjects of category $3$ are said ``coinfected" with malaria and arbovirus.

Our aim is to differentiate febrile syndroms that could be due to arbovirus from febrile syndroms that could be due to malaria. As coinfection in a single patient may change the spectrum of clinical symptoms, we want to identify those features that predict arboviral infection to improve medical and treatment diagnosis in the primary care setting.

In this study, arboviral cases are diagnosed  by the detection of IgM. We considered that an individual was positive for arboviral infection if he/she was tested positive to IgM. Ignoring individuals with missing data, we obtained a data set of size $n=12288$ (\textit{IgM data}) which is summarized in Table~\ref{tab1}. We can see that this data set is very unbalanced ($3$ arboviral or coinfected cases per $1000$ patients) and will require a specific statistical analysis.

\begin{table}[h!]
\centering
\begin{tabular}{|c|c|c|c|}
\hline 
\diagbox{\textbf{Arbovirus}}{\textbf{Malaria}} & $+$ & $-$& Total \\ 
\hline 
$+$  & $18 $ $(0.15\%)$ &$21 $ $(0.16\%)$& $39$ $(0.31\%)$ \\ 
\hline
$-$  & $7\;069 $ $(57.53\%)$& $5\;180 $ $(42.16\%)$& $12\;305 $  $(99.69\%)$ \\ 
\hline 
Total & $7\;087 $  $(57.68\%)$& $5\;201 $ $(42.32\%)$& $ 12\;288$ \\ 
\hline 
\end{tabular}
\caption{IgM data. A summary of the response variables.}
\label{tab1}
\end{table}

\begin{figure}[!h!!]
\centering
\includegraphics[height=9.68cm,width=14cm]{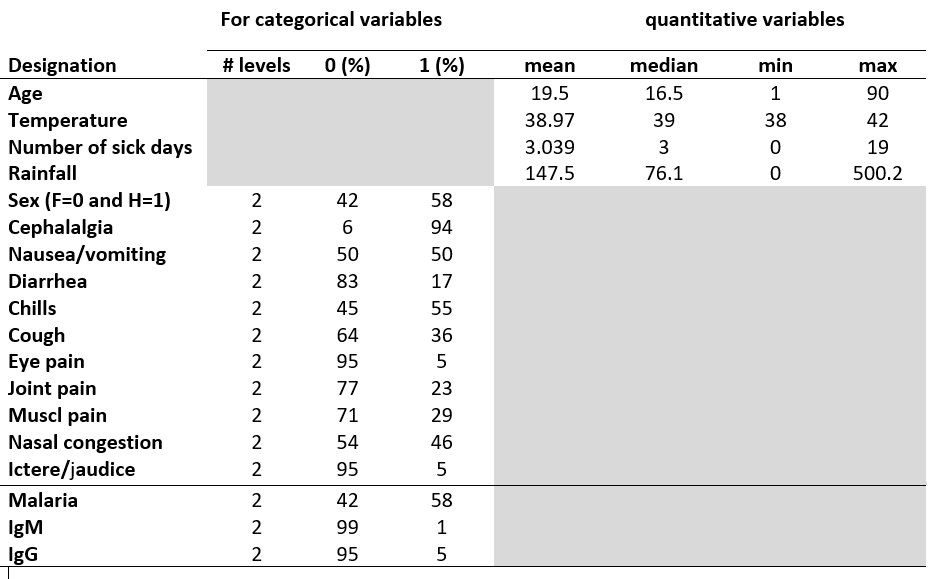}  
\caption{List of variables}
\label{fig1}
\end{figure}

In the data set, there are four quantitative covariables: the measured body temperature (in Celsius degrees), the number of sick days defined as the number of days between the date of symptoms onset and the date of consultation, the patient's age (in year) and the rainfall measure (in millimiters) which is a proxy for the season (rainy or dry). The individual rainfall measure corresponds to the rainfall measure of the patient's month of consultation. The eleven qualitative covariables are the patient's gender and ten other binary variable, which record presence or absence of ten symptoms: headache, eye pain, muscle pain, join pain, cough, nausea or vomiting, chills, diarrhea, nasal congestion and icterus and/or jaundice. All the variables of the data sets are summarized in Figure~\ref{fig1}. 
\section{Statistical analysis of the coinfection influential factors}
\label{sec3}

The objective of this section is to propose a methodology that can identify the important symptoms for the arbovirus diagnosis and can help making decision for arbovirus treatment in absence of laboratory confirmation. 

Variable selection is appreciable in medical data analysis as the diagnosis of the disease could be done on a minimum number of clinical measures. Reducing the number of relevant covariates may also produce more accurate classification results. 
In a first step, we select relevant covariates that explain the disease status typically via a multinomial logistic model. The statistical analysis is challenging because of the small number of instances of the arboviral class ($39$) with respect to the total number of observations ($12\;288$). The cases that are more important for the study are rare and few exist on the available training set. We face what is usually known as a problem of imbalanced data sets. To handle this problem, we proposed to randomly remove observations from the majority class to prevent its signal from dominating the fitting procedure. We applied to our imbalanced $IgM$ data set a common undersampling technique to obtain a more balanced data distribution. As the data distribution is changed, it is expected that the fitted models are biased to the goals of the user and are more interpretable in terms of these goals.

In a second step we investigate the robustness of the variable selection using random forests. Introduced by \cite{Breiman}, random forests (RF hereafter) are a robust nonparametric method to deal with classification problems. They require only mild conditions on the data generating model. They are also less sensitive to weaknesses in the data, because the randomized tree generation procedure ensures that all covariates are more equally evaluated. Moreover, RF decision trees often perform well on imbalanced data sets because ensemble methods offer ways to rebalance the distributions in varied ways. In this study, RF models have the advantage of providing a ranking of covariates using the RF score of variable importance that is a useful and effective tool to find important covariates for interpretation. 

In a third step, we quantify the effects of the selected covariates using odds ratios. We compute odds ratios for one disease category relative to another one and we contrast the effects of the covariates on the disease category, arboviral monoinfection, malaria monoinfection and coinfection.

\subsection{Multinomial logit model}
\label{mod1}
We recall that $Y$ is the response variable indicating the class of the disease: ``Other febrile illnesses" ($ Y=0 $), ``arboviral monoinfection" ($ Y=1 $), ``malaria monoinfection" ($ Y=2 $) and ``coinfection" ($ Y=3 $). Let $X=(1,X_1,\ldots,X_p)$ be the vector of the $p$ covariates. 
For an individual with covariates $ X=x $, we want to predict the probability of belonging to the class $k$ given $x$, 

$$
\pi_k(x)=\mathrm{P}(Y=k|X=x), \quad k=0,1,2,3.
$$
The multinomial logit model assumes the existence of $\beta_1$, $\beta_2$, $\beta_3\in \mathbb{R}^{p+1}$ such that, for each $k=1,2,3$ and each vector of covariates $x$,
\begin{eqnarray} \log \frac{\mathrm{P}(Y=k|X=x)}{\mathrm{P}(Y=0|X=x)}&=&\langle x,\beta_k\rangle
\label{eqlogit}
\end{eqnarray}
where $$\langle x,\beta_k\rangle=\sum_{j=0}^px_j\beta_{kj}$$
and $x_0=1$ to include the intercept parameters $\beta_{k0}$, $k=1,2,3$. The reference modality is class 0. \\
Consequently, for each $k=1,2,3$ and each vector of covariates $x$,
$$  \mathrm{P}(Y=k|X=x)=\frac{\exp(\langle x,\beta_k\rangle)}{1+\sum_{l=1}^3\exp(\langle x,\beta_l\rangle)}
$$
and
$$ \mathrm{P}(Y=0|X=x)=\frac{1}{1+\sum_{l=1}^3\exp(\langle x,\beta_l\rangle)}
\cdot$$
From the computation of the maximum likelihood estimates $\widehat{\beta}_k$, we derive for $k=1,2,3$,
\begin{eqnarray}
\label{eqprob}
\widehat{\pi_{k}}(x)=\frac{e^{\langle x,\widehat{\beta_k}\rangle}}{1+\sum_{l=1}^{3}e^{\langle x,\widehat{\beta_l}\rangle}}\cdot
\end{eqnarray}


\subsection{Fitting strategy for handling imbalanced $IgM$ data}

The $IgM$ data set contains $18$ arboviral monoinfection cases, $21$ coinfection cases, $5\;180$ other febrile illness cases and $7\;069$ malaria monoinfection cases. Trained on the original $IgM$ data set, the fitted logit model only predicted classes 0 and 2, which means it ignores the two minority classes 1 and 3 in favour of the majority classes. Applying resampling strategies to obtain a more balanced data sample is an effective solution to the imbalance problem (see \cite{Branco} for a survey of existing methods). Two of the most simple resampling approaches are undersampling and oversampling. Since the $IgM$ is highly imbalanced with a large number of observations in the two majority classes, we used a random undersampling strategy that removes observations and reduces the sample size.  
We sampled without replacement $50$ cases from each of the two majority classes to create a balanced sub-sample of size $18+21+50+50=139$. Trained on a sub-sample, the model predicted four classes. 

Undersampling results in loss of information and the risk of removing relevant observations is present. To overcome this problem, we repeated the sampling step a thousand times and worked with $1\;000$ balanced sub-samples of the $IgM$ data set. The multinomial model was fitted to each sub-sample and a stepwise covariate selection was performed (see Figure 5 in the supplementary material). 
The observed variability of the $1\;000$ covariate selections raised robustness questions. To answer this point, we conducted a nonparametric analysis based on the RF algorithm. In recent years, several methods involving the combination of resampling and ensemble learning have appeared in the imbalanced distributions literature (\cite{Branco}). We found that the importance score based on random forests yielded a convenient way to summarize the information obtained from the $1\;000$ sub-samples.

\subsection{Variable selection using random forests}
A random forest is an ensemble of unpruned trees, induced from bootstrap samples of the training data, that uses random covariate selection in the tree construction process. Prediction is made by aggregating the predictions of the ensemble, using the majority vote rule. 

One of the most widely used RF score of importance of a given variable is the Mean Decrease of Accuracy ($MDA$) in predictions. It is based on the out-of-bag (OOB) error. For each tree $t$ of the forest, consider the associated $OOB_t$ sample (data not included in the bootstrap sample used to construct $t$). Denote by $errOOB_t$ the misclassification rate of tree $t$ computed on this $OOB_t$ sample.  Then, randomly permute the observed values of covariate $X_j$ in $OOB_t$ to get a perturbed sample and compute $errOOB_t^j$, the error of $t$ on the perturbed sample. Variable importance of $X_j$ is then given by
$$MDA(X_j)=\frac{1}{ntree}\sum_{t=1}^{ntree}\left( errOOB_t^j- errOOB_t\right),$$
where $ntree$ denotes the number of trees of the RF.
The higher the $MDA$, the more important the variable is. Several variable selection procedures using RF are based on this quantification of variable importance. 

\noindent Using R packages, we made the following implementation choices: \verb"randomForest" for RF fitting and $MDA$ calculation, \verb"VSURF" for selecting the important variables. The main parameters of \verb"randomForest" were calibrated and set to their default values, \verb"ntree"=500 and \verb"mtry"=$\sqrt p$=3 (number of variables tried at each split of a tree of the RF). The variable selection strategy of \verb"VSURF" is based on a two-stage procedure (\cite{Genuer2}): 1. the covariates are ranked by sorting their variable importance measures in descending order and the covariates whose importance is less than a threshold (the minimum value of the standard deviations of the importance measures) are eliminated; 2. a sequence of nested models starting from the one with only the most important
variable and ending with the one involving all important
variables kept previously is considered; the variables of the model leading to the smallest $OOB$ error are selected. An advantage of using VSURF is that this procedure does not require the choice of tuning parameters.
 
\begin{figure}[h!]   
\includegraphics[height=10cm,width=12cm]{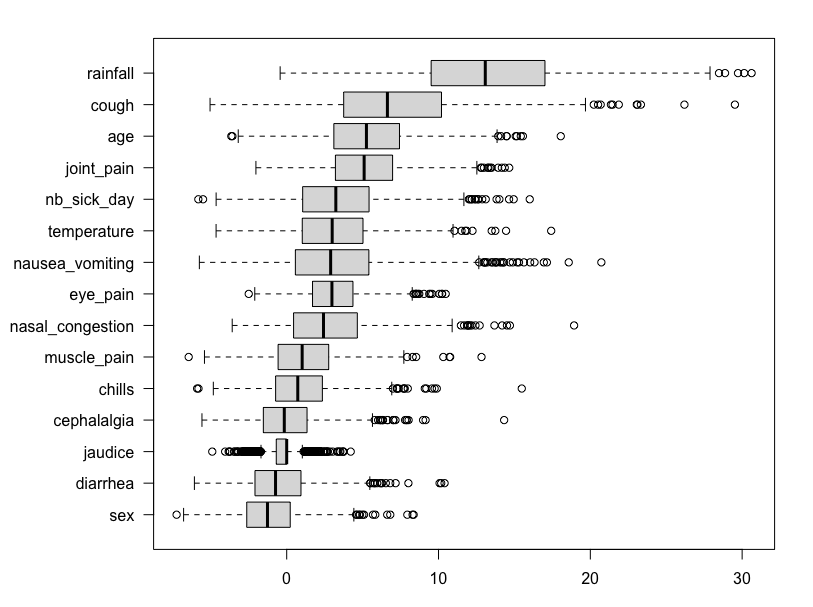}   
\caption{A variable importance plot for the \textit{IgM} data set. Each boxplot summarizes the distribution of the variable importance among $1000$ \textit{IgM} sub-samples. }
\label{fig4}
\end{figure}

\begin{figure}[h!]
\centering
\includegraphics[height=10cm,width=12cm]{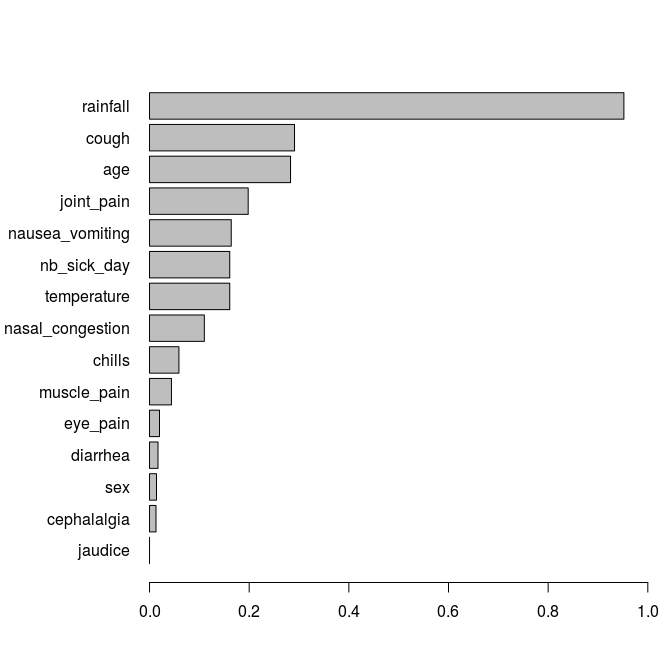}  
\caption{Ranking by VSURF: for each variable, the length of the bar corresponds to the empirical probability to be selected by VSURF among $1000$ \textit{IgM} sub-samples}
\label{fig5}
\end{figure}

%
%
%
Figure~\ref{fig4} ranks the variable importances (MDA) of the $15$ covariates across the $1\;000$ sub-samples. First, \textit{rainfall} is the most important covariate; a second group of less important covariates is formed by \textit{cough, age and joint pain}; then comes a group of five covariates: \textit{number of sick days, temperature, nausea or vomiting, eye pain and nasal congestion}; finally, six unimportant covariates are displayed: \textit{muscle pain, chills, cephalalgia, jaudice, diarrhea and sex} . The boundary between the two last groups is not clear and we used the VSURF procedure to separate the important covariates from the other ones. We can notice on the plot that both MDA level and variability are larger for relevant variables; as explained by \cite{Genuer1}, this is expected and the VSURF threshold value is based on $MDA$ standard deviation estimation.
Figure~\ref{fig5} summarizes the results of the VSURF selection procedure based on the $1\;000$ sub-samples. The covariate \textit{rainfall} ($95.2\%$) is almost always selected. Next, the more often selected variables are \textit{cough} ($29.1\%$), \textit{age} ($28.3\%$), \textit{joint pain} ($19.8\%$), \textit{nausea or vomiting} ($16.4\%$), \textit{number of sick days} ($16.1\%$), \textit{temperature} ($16.1\%$) and \textit{nasal congestion} ($11\%$), in decreasing order. The other covariates are selected in less than $10\%$ of the samples. 

We set different random seeds and we found that, for our purpose of selecting significant covariates, aggregation of $1\;000$ RF classifiers learned from $1\;000$ randomly balanced sub-samples yielded stable selected variable sets.  

\subsection{Influence of selected covariates on disease status}
\label{sec33}
In the previous sections, the RF variable importance results on the $IgM$ sub-samples produced a robust ranking of the covariates. From these results, we decided to fit multinomial model with eight covariates (\textit{age, temperature, number of sick days, rainfall, nausea or vomiting, cough, nasal congestion and joint pain}) to the data set of our analysis and to further quantify the effects of the covariates in this model. 

Within the multinomial logit model, we can quantify the effect of a variable in terms of an odds ratio or its logarithm. The odds that $Y=k$ occurs for an individual with covariates $X=x$ is the ratio of $\mathrm{P}(Y=k|X=x)$ divided by $\mathrm{P}(Y=0|X=x)$, $k=1,2,3$. Then, the log odds of category $k$ is given by Equation~(\ref{eqlogit}) : $$\log \mathrm{odds}(Y=k|X=x) = \langle x,\beta_k\rangle.$$
Thus the multinomial logit model is a linear regression model in the log odds. The parameter component $\beta_{kj}$ can be interpreted as the change in the log odds per unit change in the continuous covariate $X_j$, if all other covariates are held constant. The odds ratio (OR) of category $k$ for a $d$ units increase of $X_j$, all other covariates remaining constant, is defined as
$$OR_k (d)= \frac{\mathrm{P}(Y=k|X_j+d)/\mathrm{P}(Y=0|X_j+d)}{\mathrm{P}(Y=k|X_j)/\mathrm{P}(Y=0|X_j)}=\exp(\beta_{kj}d).$$ 
Once $\beta$ is estimated, one can estimate any odds or odds ratios. An OR equal to one means that  a change in covariate $X_j$ has no effect on the odds of category $k$; if $OR_k(d)>1$ ($OR_k(d)<1$), the effect of an increase of $X_j$ is to increase (decrease) the odds of category $k.$ The risk ratio $\mathrm{P}(Y=k|X_j+d)/\mathrm{P}(Y=k|X_j)$, which could be more interpretable in terms of predicted probabilities instead of odds, depends on the values of all other covariates. ORs are similar to risk ratios if the risk is small, otherwise ORs overestimate risk ratios. 

\begin{figure}[!h!!]
\centering
\includegraphics[height=8cm,width=4.13cm]{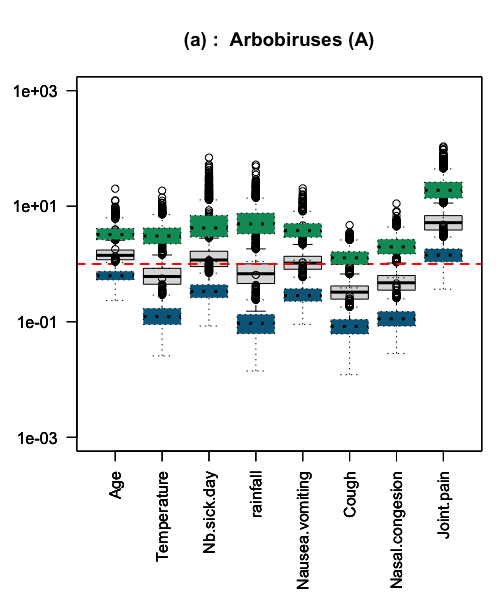}
\includegraphics[height=8cm,width=4.13cm]{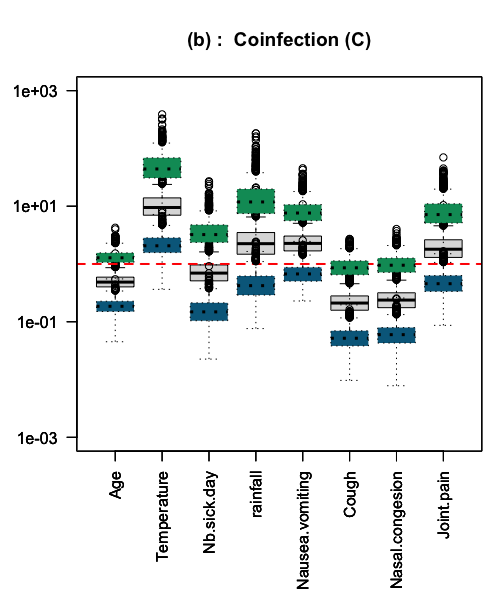} 
\includegraphics[height=8cm,width=4.13cm]{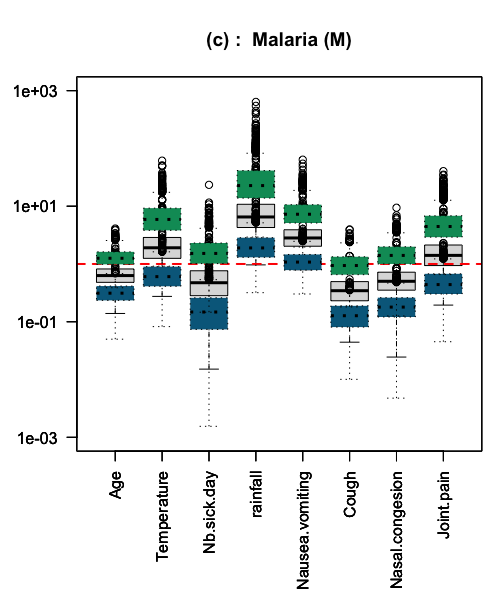}   
\caption{$IgM$ data: boxplots of 1000 odds ratios with respect to the reference category (grey full line boxplot);  boxplots of the associated confidence intervals are shown in dotted line (green for upper and blue for lower bound); (a) Arbovirus (b) Coinfection (c) Malaria.}
\label{fig7}
\end{figure}

\smallskip
For each covariate, we computed the odds ratios $OR_k$, $k=1,2,3$ and their confidence intervals for each disease. Figure~\ref{fig7}  display the OR by which the odds increases for a certain change in a covariate, holding all other covariates constant. The ORs associated with binary variables (\textit{nausea/vomiting, cough, nasal congestion and joint pain}) were computed by comparing the two modalities: $0$ for absence and $1$ for presence of the symptom. 
We computed the ORs resulting from increasing \textit{temperature} from $38$ to $40$ degrees Celsius ($d=2$) and from increasing \textit{Number of sick days} from $2$ to $6$ days ($d=4$). The outer quartiles of \textit{Age} are $8$ and $28$ years ($d=20$), so we computed the half-sample OR for age. Similarly, we computed the half-sample OR for a \textit{rainfall} of $14$ mm compared to a \textit{rainfall} of $370$ mm ($d=356$).

\begin{figure}[!h!!]
\centering
\includegraphics[height=8cm,width=4.13cm]{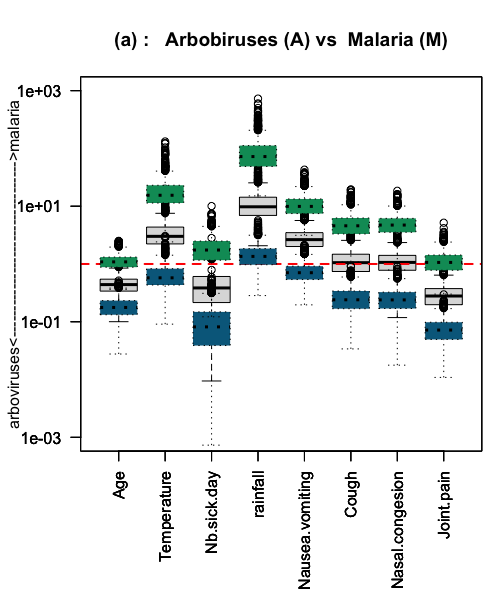}
\includegraphics[height=8cm,width=4.13cm]{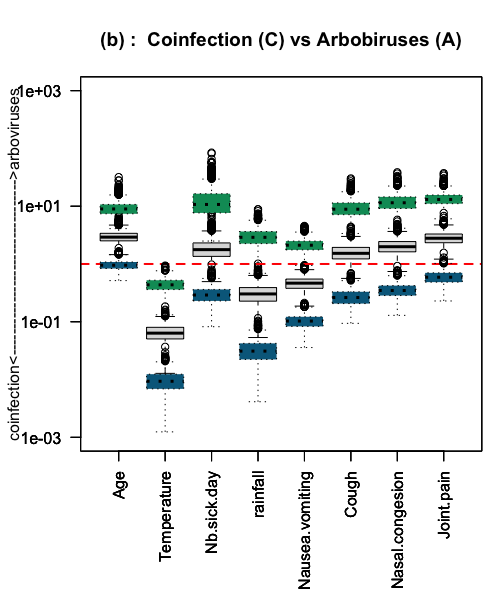} 
\includegraphics[height=8cm,width=4.13cm]{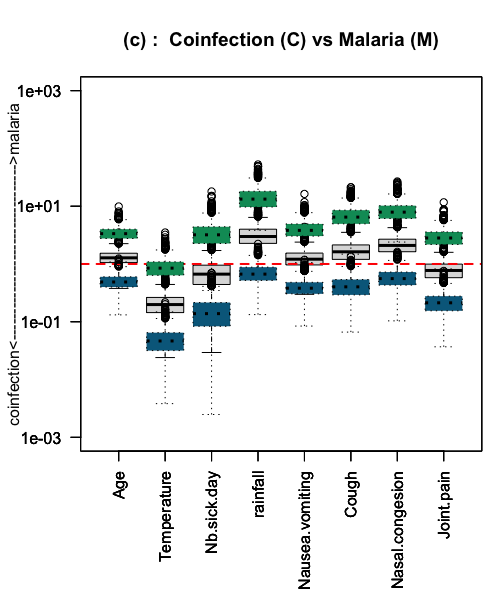}   
\caption{$IgM$ data: boxplots of 1000 odds ratios between two categories (grey full line boxplot); boxplots of the associated confidence intervals are shown in dotted line (green for upper and blue for lower bound); (a) Arbovirus \textit{vs} Malaria (b) Coinfection \textit{vs} Arbovirus  (c) Coinfection \textit{vs} Malaria.}
\label{fig8}
\end{figure}

The ORs defined previously are relative to the reference category $Y=0$. We also computed the ORs between two diseases $Y=k$ and $Y=l$ in order to 
differentiate the effect of each covariable between the three clinical groups, arbovirus vs malaria, coinfection vs arbovirus and coinfection vs malaria (Figure~\ref{fig8}):
$$OR_{k|l} (d)= \frac{\mathrm{P}(Y=k|X_j+d)/\mathrm{P}(Y=l|X_j+d)}{\mathrm{P}(Y=k|X_j)/\mathrm{P}(Y=l|X_j)}=\exp((\beta_{kj}-\beta_{lj})d).$$ 
The confidence intervals are derived from the fitted multinomial logit model and their accuracy is based on the parametric assumption that the true data generating distribution does fall in the model.

Figure~\ref{fig7} and Figure~\ref{fig8} display the sampling distribution of ORs based on the fitting of the $ 1000$ sub-samples of the $IgM$ data set. According to Figure~\ref{fig7}, we can say that rainfall and vomiting symptoms are hightly correlated with malaria monoinfections whereas joint pain is correlated with arboviral monoinfections. The odds of coinfection increases with high fever. It corroborates the conclusion of the paper \cite{sow}. Figure~\ref{fig8}(a), (b) and (c ) can be interpreted in the same way. They show that a high temperature and the presence of nausea or vomiting symptoms are mostly indicative of malaria parasite infections whereas an increase of age and of number of sick days are indicative of arboviral infections. The effects of nasal congestion and joint pain symptoms on the disease status are not clear enough to be interpreted. The main question of the study was to identify risk factors that can help doctors to diagnose a concurrent malaria and arbovirus infection. From these results, \textit{temperature} is the only risk factor that differentiates between coinfection and single infections. 

\section{Predictive analysis}
\label{sec4}
In this section we propose a methodology to discriminate arbovirus positive and arbovirus negative cases among coinfected patients.

\subsection{Testing independence between arbovirus and malaria}
\label{sec41}
In the multinomial model given by (\ref{eqlogit}) in Section \ref{mod1}, we can test the independence between arboviral and malaria infections.

\medskip\noindent 
The joint statistical distribution of arboviral infection ($Y\in \lbrace 1,3\rbrace $) and malaria infection ($Y\in \lbrace 2,3\rbrace $), is given in Table~\ref{tab3}. Independence between arboviral and malaria infections means that for all $(a,\, m) \in \lbrace 0,1\rbrace, $   $$P\left(A=a,\,M= m\right)=P\left(A=a\right)\times P\left(M=m\right) $$ which is equivalent to $$P\left(Y=2m+a|X=x\right)=P\left(Y\in\lbrace a, 2+a\rbrace|X=x\right) \times P\left(Y\in\lbrace 2m, 2m+1\rbrace|X=x\right) $$ for all $(a,\, m) \in \lbrace 0,1\rbrace, $ where 
$P\left(Y\in \lbrace 1,3\rbrace | X=x\right)$ corresponds to the probability to  belonging of categories $1$ or $3$,  $P\left(Y\in \lbrace 2,3\rbrace | X=x\right)$ corresponds to the probability of categories $2$ or $3.$
The independence hypothesis can be written in terms of parameters as:
$$H_0:\qquad ``\beta_3=\beta_1+\beta_2".$$

\medskip\noindent
The Wald statistic to test $H_0$ against its two-sided alternative is computed as
$$
W= h(\widehat{\beta})^{T}\Sigma^{-1} h(\widehat{\beta}),
$$
with $h(\widehat{\beta})=\widehat{\beta}_3-\widehat{\beta}_1-\widehat{\beta}_2$ and $ \Sigma=DVD^{T} $ where $ D=(-Id_{p+1},-Id_{p+1},Id_{p+1}) $; $Id_{p}$ is the $p\times p$ identity matrix 	and $V$ is an estimator of the variance of $ \widehat{\beta}= (\widehat{\beta}_1,\widehat{\beta}_2,\widehat{\beta}_3)^T$.
Under $H_0$, $W$ is asymptotically distributed as a chi-square variable with ($p+1$) degrees of freedom. Under $H_1,$ $W$ converges to infinity as the sample size goes to infinity. 

\begin{table}[!!!h!!!]
\begin{center}
 \begin{tabular}{|c|c|c|c|}
\hline
  & $A=0$ & $A=1$ & Law of  $M$ \\ 
\hline 
$ M=0 $ & $ \pi_0 $ & $ \pi_1 $ &$P(
M=0)=\pi_0+ \pi_1$ \\ 
\hline 
$ M=1 $& $ \pi_2 $ & $ \pi_3 $ & $P(M=1)=\pi_2+ \pi_3$ \\ 
\hline 
Law of $A$ & $P(A=0)=\pi_0+ \pi_2$  & $P(A=1)=\pi_1+ \pi_3$ & 1 \\ 
\hline 
\end{tabular} 
\end{center}
\caption{Joint distribution of arboviral infection and malaria infection}
\label{tab3}
\end{table}

\medskip
%
%

\subsection{Diagnosis of arboviral disease}
 In absence of rapid arbovirus detection tests, the aim is to provide a decision support tool to determine if an arbovirus could be responsible for the clinical symptoms of the patient coinfection.  We propose to base the diagnosis on the conditional probability to be coinfected $q(x)=P\left(Y=3|Y\in \lbrace 2,3\rbrace, X=x\right)$ given that malaria infection is observed. This probability is the quantity of interest because arboviral infections are considered by healthcare workers only if malaria tests are negative. 
 
\medskip In the previous section, it is shown that we can test the independence hypothesis between malaria and arboviral infections. If this test is rejected, then we can derive the probability $q$ to be coinfected given that malaria infection  is observed. This probability can be computed in function of the $ \pi_k $ probabilities estimated from the multinomial logit model. For $X=x$, 
$$ \hat{q}(x)=\frac{\widehat{\pi_3}(x)}{\widehat{\pi_3}(x)+\widehat{\pi_2}(x)}=\frac{e^{\langle x,\widehat{\beta_3}\rangle}}{e^{\langle x,\widehat{\beta_3}\rangle}+e^{\langle x, \widehat{\beta_2}\rangle}}\cdot$$

This probability can be used to differentiate whether the illness to be treated should be arbovirus or malaria. We propose a binary classification rule and we predict an arbovirus illness if the estimated coinfection probability is greater than a threshold value $\gamma$:
$$\left\{
\begin{array}{ cl}
\mbox{If}\,\,\, q(x)\geq \gamma:  & \mbox{\textit{arbovirus positive case},} \\ 
\mbox{If}\,\,\, q(x) < \gamma: & \mbox{\textit{arbovirus negative case}.}
\end{array}\right.
$$
The evaluation of the classification is based on the confusion matrix and the overall classification accuracy.
The confusion matrix is used to compute true arbovirus positives (TP), false arbovirus positives (FP), true negatives (TN) and false negatives (FN). A global performance measure is the miss-classification rate (MCR) defined as:
$$\textrm{MCR} = \frac{FP+FN}{N},$$ with $N=TP+FP+TN+FN.$

Our analysis is based on a real-life medical data set. In the original IgM data set, arbovirus positive individuals are identified as individuals likely to be in the early stages of arbovirus illness. It is the relevant data set for the classification problem. However, the positive cases constitute only a very small minority class of the data (39 positive cases over 12288 individuals).  Based on these data, the computation of the independence test is very sensitive to the fluctuations  of the sub-sampling procedure and the classification procedure could not be implemented. Instead, we propose a simulation study based on a balanced data set to illustrate our classification procedure. 

\subsection{Simulation study} 
\noindent\textbf{Simulated data.}  
Taking advantage of the previous influencing factors analysis (Section \ref{sec33}), we simulated data using a multinomial model similar to the one previously estimated. The eigth covariates were generated  from distributions similar to those observed in the real data set. The beta parameters values are given by Table 7 in supplementary material. They were chosen according to the conclusions of the statistical analysis of the influential factors. The larger values emphasize the influence of the associated covariates that are positively correlated to each disease category.  For example, the parameter value associated with the \textit{number of sick days } covariate is larger for the arbovirus category than for the malaria category. Based on this generative model, we computed the probabilities of belonging to each category and generated the $Y$ response to be the modality with the greatest probability. We  used this procedure to simulate a data set of size  $n=5000$ which is summarized in Table 6 and Table 8 in the supplementary material.

\smallskip
\noindent \textbf{Independence between abovirus and malaria.}
We fitted the multinomial model to the simulated data and tested the independence between malaria and arbovirus. We obtained that the independence hypothesis was rejected with a p-value equal to $1.13\times 10^{-4}.$ Then we derived the probability to be coinfected given that malaria infection is observed and performed the classification procedure.

\begin{figure}[!h!!]
\centering
\includegraphics[scale=0.3]{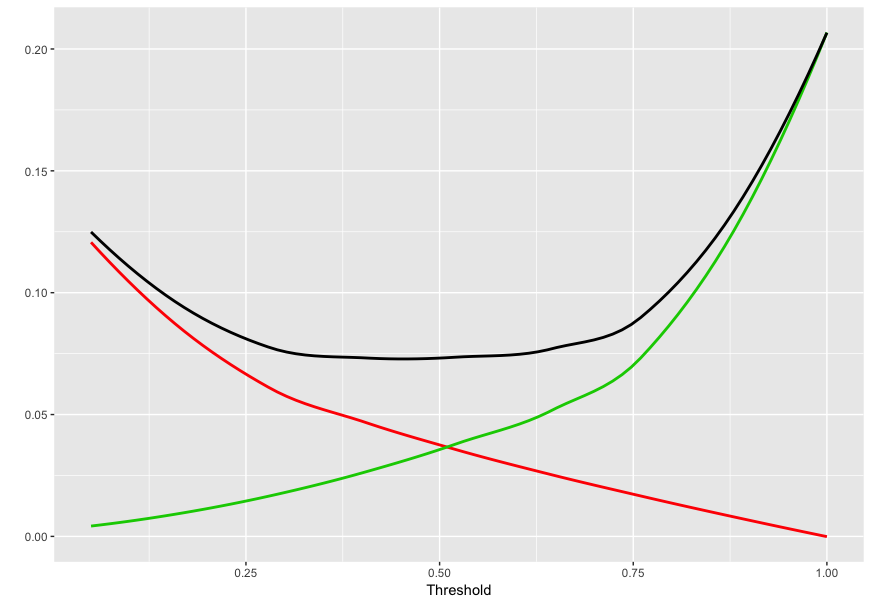}  
\caption{Cross-validation miss-classification rate. The MCR is shown in black as full line. Increasing $\gamma$ increases the number of FN (green line) and decreases the FP (red line).}
\label{fig9}
\end{figure}

\smallskip
\noindent \textbf{Diagnosis of arboviral disease.}
We randomly divided the simulated data set into two part, a training data set of size $3333$ and a test data set of size $1667$. The classification was applied only to individuals infected with malaria parasites, namely $1925$ individuals in the training data set and $626$ individuals in the test set. 
We computed the five-fold cross-validation estimator of the MCR and we chose the classification threshold value $\gamma$ as the minimizer of the MCR. We can see on Figure~\ref{fig9} that the optimal value of this threshold is $\gamma=0.45.$ Five-fold cross-validation was run several times and the optimal value of $\gamma$ was found to be quite stable. Based on this $\gamma$ value, we performed the classification to predict the type of illness that has affected the patient. Predicted and actual arbovirus cases were compared using the test set, as presented in Table~\ref{tab5}. The rows of the matrix are actual classes and the columns are the predicted classes. We observe that the corresponding test MCR is $7.83\%.$ The ROC curve of the classification is presented in Figure 7 of the supplementary material. Based on the simulated data set, the accuracy of the classification is quite good ($92.17 \%$). This suggests that this predictive analysis can be medically valuable to identify arboviral cases among coinfection cases.

\begin{table}[h!]
\begin{center}
    \begin{tabular}{|c|c|c|c|}
\hline 
\diagbox{\textbf{\textit{Actual}}}{\textbf{\textit{Predicted}}}  & $  0 $ & $ 1 $ \\ 
\hline 
$ 0 $& $421$ & $29$ \\ 
\hline 
$1$ & $ 20 $ &$156$\\ 
\hline 
\end{tabular}
\caption{Confusion table with $\gamma=0.45.$ Each row represents the instances in the actual class and each colum represents the instances in the predicted class. Class $0$ for malaria monoinfection and class $1$ for coinfection.}
\label{tab5}
\end{center}
\end{table}
\medskip\noindent

%


\section{Discussion}
\label{sec5}

\label{sec5}

Misdiagnosis of arbovirus coinfections as malaria infections may increase the spread of arbovirus diseases in areas where fast diagnostic assays are not available. This study proposes an appropriate statistical methodology that can assist doctors in the elaboration of the differential diagnosis of febrile cases for arboviruses.

To analyze coinfection data we propose a methodology with three steps: 1. a variable selection with random forests; 2. an analysis of the influent factors through multinomial model fitting and odd ratios computation; 3. a predictive analysis based on coinfection probabilities.  From our experiments, we can say that the random forests algorithm is a robust method to select the important variables for the different diseases. The analysis of the odd ratios allows to identify the risk factors that characterize each disease. We observed that higher values of number of sick days and of age are mostly indicative of arboviral disease while higher values of temperature and presence of nausea or vomiting symptoms during the rainy season are mostly indicative of malaria disease. The results also pointed out that a high-grade fever could be considered as a differential diagnostic for malaria and arbovirus coinfection, which is in agreement with the study of \cite{sow}. The proposed predictive analysis was illustrated on a simulated data set. We show that using data with enough signal, we can identify coinfected patients to be treated for arbovirus with great accuracy. A future study will apply this methodology to coinfection data between viral and bacterial infections collected in Senegal by Institut Pasteur de Dakar from 2015 to 2017.

\bigskip\noindent


\section*{References}
\bibliographystyle{elsarticle-num}
\bibliography{arbo_palu.bib}

\end{document}